# Onshore to Near-Shore Outsourcing Transitions: Unpacking Tensions

Bilal Raza, Tony Clear and Stephen G. MacDonell
*SERL, School of Computer & Mathematical Sciences*
*Auckland University of Technology*
*Auckland, New Zealand*
braza@aut.ac.nz, tclear@aut.ac.nz, smacdone@aut.ac.nz

**Abstract**

*This study is directed towards highlighting tensions of incoming and outgoing vendors during outsourcing in a near-shore context. Incoming-and-outgoing of vendors generate a complex form of relationship in which the participating organizations cooperate and compete simultaneously. It is of great importance to develop knowledge about this kind of relationship typically in the current GSE-related multi-sourcing environment. We carried out a longitudinal case study and utilized data from the 'Novopay' project, which is available in the public domain. This project involved an outgoing New Zealand based vendor and incoming Australian based vendor. The results show that the demand for the same human resources, dependency upon cooperation and collaboration between vendors, reliance on each other system's configurations and utilizing similar strategies by the client, which worked for the previous vendor, generated a set of tensions which needed to be continuously managed throughout the project.*

## 1. INTRODUCTION

Globalization and advances in telecommunications mean that organizations have access to a range of software sourcing strategies that can span geographical boundaries, enabling them to expand their customer base, employ new resources and work more closely with new, previously out-of-reach customers [1]. Outsourcing is one such strategy [2] in which a client company contracts out part or all of their software development activities to an external vendor company [3]. Although outsourcing has led to many benefits it has also brought about adverse consequences, such as the loss of internal capabilities and skills at the hands of vendor organizations [4]. As with all sourcing strategies, outsourcing is not inherently good or bad; rather, the outcome is determined by how well it is managed before and after a contract is signed [5]. While outsourcing has become a 'business as usual' strategy it has been noted that about half of such contracts are discontinued either in order to switch/transition to a new service provider or to bring the previously outsourced activities in house – the latter being referred to as Backsourcing [6]. So what happens when such contracts are terminated? There is limited research that deals with such questions [7]. In this paper we report the results of an in-depth longitudinal case study that we carried out to understand the process of vendor switching. The study is focused around a project in which the main client switched its payroll system and associated services from an onshore to a near-shore service provider. In this paper we focus in particular on the tensions that arose for and between the incoming and outgoing vendors, in a context in which the vendors had to cooperate and compete at the same time.

## 2. CASE INTRODUCTION

The widely publicized 'Novopay' project is an ongoing New Zealand-wide transition from an onshore based service provider, Datacom, to an Australian based nearshore provider, Talent2. The project centered on the Ministry of Education's (MoE-New Zealand) payroll processing system along with its service model, which was responsible for the payment of teachers and education sector staff. The project involved diverse range of stakeholders and included the setting up of a new service model following refined business processes. Although, a core function in organizations, payroll is rather unforgiving and critical. Even minor problems can blowout into significant issues for employees and their representatives. Since its roll-out Novopay has faced numerous issues relating to the reliability and accuracy of payruns. The inability to address these issues in a timely manner has resulted in major sector frustration and a strongly negative public perception.

## 3. RESEARCH METHOD

### A. Case Study

Successful implementation of large scale software and IT-related projects are frequently derailed by non-technical challenges due to organizational issues, human behavior and poor project management [8]. Given the complex context of this research and the central role of change over time, a longitudinal case study was undertaken, enabling us to focus upon events as they unfolded. By following such an approach [9] we were better able to gain deeper understanding about why events occurred and the threads to be followed. Case study research satisfies a desire to understand complex social phenomenon through focus on a specific 'case' while retaining a holistic and real-world perspective [10]. Due to the relevance of the case study methodology to understanding a phenomenon in its natural setting we chose to use it as the main methodology of this research.

Table I. Data Sources

| Internal Reports and Meeting Minutes | Communication |
|---|---|
| Memos of Ministry of Education (MoE) | Correspondence between representatives of MoE and Exec. of the main vendor |
| Cabinet Meeting Minutes | Correspondence emails of end users |
| Status Reports | **Project Related Documents** |
| Steering Committee Meeting Minutes | Risk Registers |
| Payroll Reference Group Meeting Minutes | Project Initiation Doc. |
| Novopay Board Meeting Minutes | Request For Proposal Doc.- including revised versions |
| Quarterly Meeting Reports for High Risk Projects | Business Case Doc. |
| **Periodic Reports of External Companies** | Fallback plans and proposals |
| PwC | Test Plans and strategies |
| Deloitte | Communication plans |
| Equinox | Reports about variations in the agreement |
| Maven | Progress review reports |
| IQANZ | Surveys carried out from end-users |
| Extrinsic | Remedial plans and programs |

**B. Data Management and Analysis**

The availability of documents and reports relating to public sector projects may lend opportunities for researchers to utilize this empirical evidence to build knowledge, and to provide guidelines and archetypes for other similar projects. In accordance with the Official Information Act 1982 of New Zealand, documents and reports about the high profile and troubled Novopay project were released to the public in 2013 - we took this as an opportunity and carried out a pilot study [11] by analyzing a subset of the data. After receiving positive feedback we embarked upon full-scale analysis of all the reports and documents. The results reported here are based upon this analysis. The data source files were published on MoE website and can be accessed through this web-link:
www.minedu.govt.nz/theMinistry/NovopayProject.aspx.

Table I lists these sources. The data analysis process followed an iterative strategy. During the first stage of 'key-point extraction' we analyzed and extracted sections of text if they addressed one or more of the following: *Actions performed by stakeholders; if possible, consequences of those actions taken; issues faced by stakeholders; decisions taken by stakeholders; concerns shown by stakeholders; Risks and issues considered for the project.* These criteria were chosen to highlight the candidate key points which could be potentially used to extract tensions.

The extracted key-points were placed in a spreadsheet for further analysis. It took more than one year for the first author to go through all the documents and extract relevant key-points. During the second stage of analysis the previously extracted key- points were coded in NVivo 10. This stage followed complementary cycles of inductive and deductive analyses. The coding process was iterative in which the initial codes were renamed, merged and/or subdivided into other codes. Throughout we ensured that the original 'chain of data' was kept with the source files. Afterwards, different codes and associated key- points were merged together in a categorizing phase. These categorized themes were then further analyzed to synthesize the tensions related to the incoming and outgoing vendor.

## 4. FINDINGS AND DISCUSSION

**A. Resource Dependency**
During the course of transition-related projects there is often a possibility to overlap resource requirements between the existing and the new vendor which may make the transition process difficult to manage. A significant risk in the effective transition related to service is how to balance business continuity until transition is complete, at the same time build capability of the new service centers prior to transition. In the Novopay case, there was also a requirement of large overlap in resources between the existing Datacom pay centers and the new Talent2 service centers. This tension risked both vendors and ultimately the client to suffer as a result. Initially, it was expected that a new vendor would seek to subcontract services from the existing operators and it was considered unlikely that the current sub-contractors would obstruct a change in prime vendor. Talent2 was expecting to deliver new services by establishing a service center in conjunction with Datacom's subcontractors and recruit staff who were currently engaged in the current payroll arrangements. Their aim was to potentially engage existing employees having relevant expertise. As the new service was expected to reduce the head- count of the service-center staff and this may have resulted in revenue reduction for sub-contractors.

The commercial negotiations and sensitivities complicated the planning and those employees were not readily available as required by Talent2. This resulted in delays and effected relationships between the incoming and outgoing vendor. One of the suggestions was to utilize the best portions of Talent2 and Datacom business continuity to deliver a Novo-Datapay solution. Although it would have been promising for MoE, both the vendors independently stated that they would not work with each other as partners. Availability of internal personnel resources to clarify business requirements are crucial for the timely completion of a project. It was well established and documented in the project initiation document that the client would ensure the availability of such personnel. Following is an excerpt taken from the Initiation Document of Novopay: "The Ministry acknowledges that in order for the Supplier to meet its obligations the Ministry will need to make relevant Ministry personnel available in a reasonably timely fashion to answer queries from the Supplier on the Ministry's business requirements and those business processes of the Ministry that will interface with the Supplier business processes to be developed" (Manage Service Delivery Service Schedule, 2008). These personnel resources were not readily available when required which resulted in delayed sign off of certain deliverables.

### B. Contrasting Culture and Lack of Cooperation

Transition-related projects often require multiple stakeholders to collaborate with each other, however, they may have contrasting and competing interests. Therefore, it is imperative for such projects to have absolute certainty about the vision, which should be communicated across all stakeholders. During the course of this project there was inefficient flow of communication between the project staff and senior leaders. It was not unanimously communicated and agreed if the project included a "black-box", or if it included a service-model as well. *"The structure of the Novopay project, where each organization is responsible for specific deliverables, with limited interaction and visibility across the work-streams, has constrained the ability of the teams to work collectively to understand the wider implications of decisions made during the development of any one deliverable. We are aware of multiple opinions across the project teams regarding barriers to communication within the project and hence a lack of a common understanding of the challenges being faced" (Novopay Review Report by Extrinsic, Jan 2010).*

This resulted in lack of trust, fatigue, disrespectful behaviors, blame, non-delivery, little celebration and a sense of not being listened to. Two permanent Subject Matter Experts decided not to extend their contracts beyond the current expiry date. They expressed concerns about the existence of a 'blame' culture, and felt under-valued. Concerns were also raised about the bureaucratic culture of MoE. *"It was noted that there was difficulty in getting consistent resources out of the MOE [Ministry] communications area. There are too many people involved with tasks being moved from one person to the next which causes delays in delivery" (Novopay Board meeting minutes, Jan2009).* Lack of teamness also effected the quality assurance activities *"The defect management system has not been robust, and the true status of deliverables has not been visible to all parties. This has all contributed to a "them and us" approach that has impeded agreement on quality criteria, a mutual understanding of the overall testing process, and what tests are required at each step of the process"(Novopay Review Report by Extrinsic, Jan 2010).* The Ministry and Talent2 also had different terminologies related to testing and there were disagreements regarding what type of testing is required at which point in time. It also indicated an underlying difference in testing philosophy — namely merely proving the service worked correctly (a requirement for contractual completion) as opposed to definitively verifying that the system and service would not fail during operational use (a longer-term business requirement for the Ministry). This created disagreements between Talent2 and Ministry regarding the scope and nature of testing and concerns were raised by Talent2 that the current testing approach would not permit the delivery of the project on time.

On the other hand, the Ministry began reviewing a sub-set of completed test scripts which were passed by Talent2 to ensure the completeness of system testing. This activity revealed that these scripts were not auditable because of insufficient detail in the desired results. Good testing practice recommends that at a minimum the expected results of a test should be recorded with the actual results noted. This was not followed in the completed test scripts reviewed by the Ministry. The Ministry therefore had lost confidence that Talent2 was appropriately testing the system.

### C. Data Management Dependency

Another area of concern during the transition project was the management of data, including its conversion from the old to the new system. In this project, this activity had key dependencies on other activities. It influenced the configuration of the system, interfaces, processes, service center training and end-user training. Talent2 showed concern that: *"they cannot control the level of data cleansing and the subsequent unknown impact that any remaining dirty data would have on dependent activities (such as testing)" (Independent Quality Assurance Summary Report, Aug2009).*

The main issue about the data was that it was held in seven separate databases having the same reference data but with different key identifiers. Each database contained data relating to different pay centers and archive storage. Data could be transferred from one database to another but this process created new records, with new reference numbers. A complete picture of a single data set could only be obtained by mapping all multiple identifiers together and 'stitching' the data together. Complaints were made by the new vendor about delayed response from the old vendor regarding requests for data, technical information and meetings. It was expressed that this may have possible effects on downstream work and completion timelines. The old vendor's response on the other hand was *"no resources available to support this work on a dedicated basis" Novopay Project Status Report Weekly, Jan 2012.* During the course of this project it was found that *"data conversion process may lead to errors, omissions and/or inaccuracies" (Novopay project risk register- collated).* Although the ownership of erroneous data lay with the Ministry and Datacom, Talent 2 did not have the ability to make decisions about the correctness of data. However Talent2 was to an extent held responsible for managing the resolution of any data issues. Datacom's point of view about the data was that it reflected the data provided by the schools so it was up them to address with their staff - although, the perception of school-staff was that it was a Ministry problem. In order to resolve these data related issues, the Ministry and Datacom contacted Schools for missing information. This activity identified instances of poor human resources practices in schools. To resolve these issues the project was tasked to negotiate the collective agreements with sector unions — to ensure that the payroll changeover had minimal impact on collective agreement negotiations and vice versa. This included an extra unplanned activity within the scope of the project. The incumbent system had been in operation since 1996.

The initial contract was set to expire in 2007, which was subsequently extended multiple times till the end of August 2012. During the course of this time, the system had been modified to fit the Ministry needs to an extent that it became an independent product having its own code-base and characteristics. The base product was written in an obsolete COBOL language for which skilled developers were difficult to find. 'On-the-fly-add-ons' which were made to enhance the functionality were not integrated into the base application and were not written and documented robustly enough. This was the situation before the transition project began and it was expected that in the Novopay change project there would be efficient processes around the *Change Control Management (CCM)* to avoid such a situation from happening again. However it was found that, *"there is an issue that changes are made in TM4 [incumbent system] in the pay period prior to go-live without prior warning to the project or change control" (Novopay project risk register – collated).*

The ripple effect of these changes were: *"problems have been identified when testing the ALESCO [new] system against converted data. Changes made in data conversion as a result of data configurations in the existing Datacom system may require changes to the configuration of the ALESCO system. Any such configuration changes will require regression testing to ensure that there are no unforeseen impacts on the payroll functionality" (Novopay Review Report by Extrinsic, Jan 2010).* Talent2 had a good base product but it lacked the required technical ability to customize and deploy their product according to the Ministry's needs.

### D. Contractual Arrangements With a New Vendor

The initiation of a new client-vendor relationship should be progressive and evolutionary, especially, in a case where the client previously established a trust-based 'Network'-like relationship with the old vendor. Expecting similar arrangements with a new vendor from the beginning may result in unexpected outcomes. Mirani in [12] argued that client-vendor relationships typically begins with a 'Contract'-based relationship in which simple applications are contracted out to the vendors. Over time more complex applications are assigned to selected vendors, which may demand an establishment of a trust-based 'Network'- like relationship. The initial approach taken in the Novopay project was to backsource the overall management to MoE and utilize services of multiple vendors. Subsequently, MoE re-evaluated its requirements and switched to a Business Process Outsourcing (BPO) based contract. A single prime vendor was chosen to manage the whole service and operation. In order to manage this BPO agreement, contracts were constructed to provide strong incentives for the quality of service to be achieved. The contract would thus considered to be the key means for deliverables. However, it may be argued that management of a large-scale transition-based BPO project through a new vendor is not a feasible policy.

Following are sample excerpts to elaborate this: *"[Ministry] acknowledge[s] that T2 has found it more difficult and more expensive to deliver on its contract commitments than it originally estimated, but this doesn't mean the Ministry must accept a higher price or reduced scope" (Executive correspondence, letter from Ministry to Talent2, May2011). "[MoE] understand your team has suggested that, if the Ministry [MoE] is not willing to provide the additional funding, the project will need to be de-scoped. It is not clear to me why the Ministry would agree to this, given that T2 has been contracted to deliver the existing project scope for an agreed fixed price" (Executive correspondence, letter from Ministry to Talent2, May2011).*

### E. Refining Business Processes

Switching of vendors and/or system-transition may provide an opportunity and temptation for the client to improve and standardize their internal business processes. This was the case in Novopay project as well which included changing of systems, switching of vendors, and service-model and refinement of business processes – all together in a single project. Introduction of such changes should be phased-in incrementally, so as to cause minimal disruption. Otherwise, as we found out from this case, the end-users may find it difficult and conflicting to manage with ease.

In the Novopay project, previously, the end-users had a dedicated pay clerk who understood the represented schools' situation and was their specific contact person. The newer service-model did not provide such customized, hands-on and highly personalized service. The previous vendor also helped end-users on one-to-one based relationships and further provided advice, guidance, interpretation and in some cases HR advice. In the new service model, this knowledge was built up at the end- users side. The main purpose of this fundamental change was to create consistency and standardization between end-users regarding collective agreement interpretations, rules and policies. For this to have worked new competencies, new technologies, new processes and new behaviors were to be developed. The underlying complexity and magnitude of the scale was not well understood and online-training was considered as an efficient method to prepare for this change. End-users were provided with direct access to their information which was expected to empower them, resulting in better service. Errors relating to data- input were prevented through validation routines implemented inside the software. However, enforcing automated-rules in the system and adding extra responsibilities to end-users resulted in resistance. They felt threatened, uncomfortable, concerned and emotional about these impositions.

Following is an excerpt taken from a Survey *"The Beta has errors, does not show allowances for staff - has fields to fill in that are new and we do not understand what they mean - you cannot run a report to check what you have entered into the system before posting and validating the info. I feel that as a user the beta is unreliable - as every time I enter data into the system - it shows up errors" (Novopay Change Readiness and Engagement Survey, Aug2012).*

In hind-sight we argue that refinement of business processes and *end-user awareness* of *transition* should be treated separately, but related projects, with careful consideration towards addressing end-user concerns within a wider change program.

### F. Single-Phase Rollout

A transition-project of this scale should be tested initially with smaller subsets of data to increase the confidence of end-users. The initial strategy of Novopay was to rollout the project in two phases. Initially South Island followed up by North Island after three months. During the course of time major concerns and risks were raised. At the South Island cutover, Datacom would still be processing the North Island payroll. Before the North Island cutover, its data would need to be converted from the Datacom system and integrated with the South Island data in the Novopay. Talent2 would then operate the North and South Island data using the Novopay system and the Datacom system would be decommissioned. However, it was noted that North Island data in the three month interim period may introduce further system or configuration changes. These changes would not have been applied to the "live" South Island data, so any integration of the two data sets had the potential to introduce further errors.

This type of staggered implementation was expected to add further complexity. Moreover, a pilot project involving 30 schools was also cancelled and replaced by a Beta Environment to be used while training end-users. Concerns were shown about: *"Staff work at multiple schools and receive a single payslip. If a pilot occurs at a selection of schools [then it] need to manage the staff moving between the pilot and other payroll. Very complex to manage moving staff between payrolls and still providing single payslip" (Novopay Project Briefing, Nov2012).* Contrasting with a view that "This big bang approach is a recipe of failure"[13]. Faults and errors are less costly to be found and fixed in the earlier stages than in the latter stages[14]. In this project, certain validation activities were pushed to the latter stages and eventually squeezed further to fit the time constraints.

Table II. SOURCES OF TENSIONS IN A TRANSITION PROJECT

| |
|---|
| Balancing the ongoing service and operations with a new setup during the interim-period |
| Dependency and reliance of human resources between the incoming and outgoing vendors |
| Dependency upon the old vendor to help interpret data stored in their system |
| Failure to negotiate with the third party contractors to support the new setup |
| Changes made to the old system, during the interim-period, resulted in data inconsistencies in the new system |
| Bureaucratic culture at MoE resulted in delays for the new vendor |
| Different terminologies causing confusion and delays |
| Management of a large-scale and complex project through a contract |
| Aiming to refine processes during switching of vendors and software system |
| Single phase roll out strategy due to complexity associated with phased transition |

## 5. VALIDITY THREATS

In this research, we relied upon the secondary data which was not gathered for the purpose of carrying out this empirical research. We had no control over this process and this can be considered as a validity threat. However, we used multiple sources of information as mentioned in TABLE I to cover the viewpoints of different stakeholders. This reduced biasness towards specific stakeholder groups and provided triangulation and saturation of themes. Furthermore, other researchers have considered secondary data to be more objective than perception-based data collected through Interviews or Surveys [15]. Qualitative studies are often criticized for the perceived difficulties in replicating them since identical circumstances may not be recreated [16].

The significance of using secondary data is that it is easily reproducible which enable other researchers to extend the original work [15] or possibly replicate the study independently. Although most of the analysis was done by the main researcher, a considerable amount of time was spent during this exercise. Regular meetings were held with the other two senior researchers from time to time to discuss the emerging themes and take feedback upon them. In order to understand the relevance of this data, we carried out a pilot study [11] in early 2013. It was based upon a subset of data which provided us confidence in the relevance of this data to describe the phenomenon of transition. To follow this up, we embarked upon a full-scale longitudinal case study. Government agencies and Ministries are often involved in software-related transition projects and, although, we don't claim that the results of this study would be widely generalizable but it will provide insights and rich description about the transition phenomenon which, if interpreted with caution, provide new avenues of risk-management in such projects. Interestingly, *Queensland (Australia) Health, Payroll Transition Project* demonstrates surprising similarities with Novopay – some of them are listed in [17].

## 6. CONCLUSION

Change is multifaceted and understanding such a complex phenomenon requires appreciation of conflicting realities, objectives and behaviors [18]. In this study we investigated the tensions of the transition process. Tensions are the outcomes of multiple, often competing, conditions which complicate the decision making. The implications of this research are that as outsourcing continues, switching from one vendor to another is also becoming common, and it is therefore important that organizations develop their readiness-models, based upon tensions, to cope with them. Traditional theory construction methodologies could lead to the development of theories which might be internally consistent but are of limited scope, which puts relatively little attention on the opportunities offered by deeper analysis of tensions and contradictions [19]. We used an alternative approach to theory building which was to look for tensions or oppositions and use them afterwards to build more encompassing theories.

Moreover, the results derived through a private organization's context may not scale and generalize to public sector projects due to differences in culture, policies, environments, governance structures etc. This warrants the creation of parallel body of knowledge related to public sector projects and in doing so the results will be reported with greater rigor. Finally, we demonstrated the value of publicly available secondary data sources for carrying out longitudinal studies. Results, from the vendors' side, show that during the interim period the incoming and going vendors have to cooperate and compete at the same time. The former was reliant upon the latter to help interpret data, system configurations and transfer of human resources. From the client side, they took vendor-transition as an opportunity to make further changes in their policies and approach – they chose a contract-based relationship, enhanced business processes and went for a big-bang approach to roll out the project. We have presented research work in progress and in the future we intend to build upon these results and include a wider set of tensions and their association with each other. In doing so we shall follow a 'process based' [20] research model and focus upon sequences of events to explain how and why particular outcomes were reached.

## REFERENCES

[1] S. Beecham, J. Noll, I. Richardson, and D. Dhungana, "A Decision Support System for Global Software Development," In 6th International Conference on Global Software Engineering Workshop, (Helsinki, 15-18 Aug.), pp. 48–53, 2011.

[2] D. Šmite, C. Wohlin, Z. Galvņa, and R. Prikladnicki, "An empirically based terminology and taxonomy for global software engineering," Empirical Software Engineering, vol. 19, no. 1, pp. 105–153, Feb. 2014.

[3] M. Ali Babar, J. M. Verner, and P. T. Nguyen, "Establishing and maintaining trust in software outsourcing relationships: An empirical investigation," Journal of Systems and Software, vol. 80, no. 9, pp. 1438–1449, Sep. 2007.

[4] M. Alaranta and S. L. Jarvenpaa, "Changing IT Providers in Public Sector Outsourcing: Managing the Loss of Experiential Knowledge," in 43rd Hawaii International Conference on System Sciences, (Honolulu,HI, 5-8 Jan.), pp. 1-10, 2010.

[5] S. Cullen, P. Seddon, and L. P. Willcocks, "Department of Information Systems London School of Economics and Political Science Working Paper Series Managing Outsourcing : The Lifecycle Imperative," MIS Quarterly Executive, vol. 4, no. 1, pp. 229-246, 2005.

[6] D. Whitten, S. Chakrabarty, and R. Wakefield, "The strategic choice to continue outsourcing, switch vendors, or backsource: Do switching costs matter?," Information & Management, vol. 47, no. 3, pp. 167–175, Apr. 2010.

[7] H. T. Barney, G. C. Low, and A. Aurum, "The Morning After: What Happens When Outsourcing Relationships End?," in Information Systems Development, G. A. Papadopoulos, W. Wojtkowski, G. Wojtkowski, S. Wrycza, and J. Zupancic, Eds., Springer US, pp. 637– 644, 2010.

[8] J. M. Verner and L. M. Abdullah, "Exploratory case study research: Outsourced project failure," Information and Software Technology, vol. 54, no. 8, pp. 866–886, Aug. 2012.

[9] J. M. Verner, J. Sampson, V. Tosic, N. A. Abu Bakar, and B. A. Kitchenham, "Guidelines for industrially-based multiple case studies in software engineering," in Proceedings of the 3rd International Conference on Research Challenges in Information Science (Fez,Morocco, 22-24 Apr.), pp. 313–324, 2009.

[10] Robert K. Yin, Case Study Research: Design and Methods, 5th ed. Sage Publications, 2014.

[11] T. Clear, B. Raza, and S. G. MacDonell, "A Critical Evaluation of Failure in a Nearshore Outsourcing Project What dilemma analysis can tell us," in 8th International Conference on Global Software Engineering (26-29 Aug., Bari), pp.178-187, 2013.

[12] R. Mirani, "Client-Vendor Relationships in Offshore Applications Development : An Evolutionary Framework," Information Resources Management Journal, vol. 19, no. 4, pp. 72–86, 2006.

[13] J. Gill, "Some New Zealand public sector outsourcing experiences Some New Zealand public sector," Journal of Change Management, vol. 1, no. 3, pp. 280–291, 2000.

[14] B. Boehm and V. R. Basili, "Software Defect Reduction Top 10 List," Computer, vol. 34, no. 1, pp. 135–137, 2001.

[15] S. C. Srivastava, T. S. H. Teo, and P. S. Mohapatra, "Business-Related Determinants of Offshoring Intensity," Information Resource Management Journal, vol. 21, no. 1, pp. 44–58, 2008.

[16] R. B. Svensson, A. Aurum, B. Paech, T. Gorschek and D. Sharma, "Software Architecture as a Means of Communication in a Globally Distributed Software Development Context," Product-Focused Software Process Improvement, Lecture Notes in Computer Science, Vol. 7343, pp.175-189, 2012.

[17] B. Raza, T. Clear, and S. G. Macdonell, "Lessons from Novopay and Queensland Health Payroll," IITP Techblog, 2014. Available: techblog.nz/677-LessonsfromNovopayandQueenslandHealthPayroll

[18] A. M. Pettigrew, "Longitudinal field research on change: theory and practice," Organizational Science, Special Issue: Longitudinal Field Research Methods for Studying Processes of Organizational Change, vol. 1, no. 3, pp. 267–292, 1990.

[19] M. Poole and A. Van de Ven, "Using Paradox to Build Management and Organization Theories," The Academy of Management Review, vol. 14, no. 4, pp. 562–578, 1989.

[20] M. Newman and D. Robey, "A Social Process Model of User-Analyst Relationships," MIS Quarterly, vol. 16, no. 2, pp. 249–266, 1992.